\documentclass[a4paper,10pt]{article} \usepackage[T1,T2A]{fontenc} \usepackage[utf8]{inputenc} \usepackage{amsmath}\usepackage{amssymb}\usepackage{amsfonts}

\usepackage{graphicx}

\def\tr{\mbox{tr}}

\usepackage[square,numbers,sort&compress]{natbib} 

\usepackage[pagebackref=true,colorlinks=true,citecolor=blue,linkcolor=blue,urlcolor = blue]{hyperref}

\begin{document} \sloppypar

\hfill UDC: 537.631, 577.352, 530.145, 57.086 \bigskip\bigskip

\begin{center} {\Large Decoupling spin relaxation and chemical kinetics in radical pair magnetism: A master equation study} \bigskip

Vladimir N. Binhi\footnote{vnbin@mail.ru} \medskip

\textit{Prokhorov General Physics Institute of the Russian Academy of Sciences\\ 38 Vavilov St., Moscow, Russian Federation} \bigskip

\end{center}

\begin{abstract} A leading theory for the biological effects of low-intensity magnetic fields is the spin-chemical Radical Pair Mechanism. This mechanism is best described by the master equation for the density matrix of an open quantum system, which includes terms for Hamiltonian evolution, thermal spin relaxation, and chemical kinetics. In this study, we have found a solution to this equation for a simplified radical pair model and shown that a significant magnetic effect occurs when the following condition is met: $\gamma H \tau \gtrsim 1 + \kappa \tau$, where $\gamma$ is the gyromagnetic ratio of the electron, $H$ is the magnetic field strength, $\tau$ is the relaxation time, and $\kappa$ is the rate of chemical kinetics. \bigskip

 \medskip \end{abstract}

\bigskip\medskip

Like a compass needle, the spin state of radical pairs in proteins can be affected by the magnetic field (MF) \citep [e.g.,][] {Buchachenko-ea-2002, Maeda-ea-2012}. This is thought to be the quantum basis for numerous observed magnetic effects in organisms \citep{Binhi-and-Rubin-2023e}. The radical pair mechanism (RPM) \citep [e.g.,][] {Zeldovich-ea-1988-e, Hore-2025} describes an intermediate complex formed in certain chemical reactions. This complex consists of a spin-correlated radical pair $\dot{\textrm{A}}\dot{\textrm{B}}$, existing in a quantum superposition of singlet $(\dot{\textrm{A}}\dot{\textrm{B}} )^{\textrm{S}}$ and triplet $(\dot{\textrm{A}} \dot{\textrm{B}} )^{\textrm{T}}$ states. In such spin-chemical reactions, MFs --- whether external or internal --- induce singlet-triplet transitions, altering the relative populations of singlet and triplet states. Consequently, the product yields of these spin-selective reactions become MF-dependent. This MF dependence underpins the RPM role in spin chemistry and biological magneto-responsiveness.

In the absence of chemical kinetics and with some idealizations formulated below, the equation for a spin system in a dimensionless form is \begin{equation} \label{mequation}  \partial_t\, \rho = - i [{\hat H},\rho] - g (\rho-\rho_{\infty }) \end{equation} where $\rho$ is the density matrix, $\partial_t$ is the time derivative operator, $\hat H$ is the Hamiltonian, $g$ is the thermal relaxation rate due to interaction with a thermostat, and $[\,,]$ denotes the commutator. We assume that thermal fluctuations are much faster than the spin dynamics. Therefore, it is reasonable to approximate the action of the thermal bath as a series of short pulses, between which the system evolves undisturbed. Then eq.~(\ref{mequation}) is a special case of the Lindblad equation for an open quantum system, with the action of the bath described by phenomenological dissipator $-g (\rho-\rho_{\infty })$ \citep [][p.~64] {Il'inskii-ea-1994}. In a weak MF, the Zeeman splitting is much smaller than $k_{\textrm{B}}T$, so $\rho_{\infty}$ is the density matrix of a fully mixed state.

The chemical kinetics in the RPM is typically modeled using an anticommutator $-\frac12\left\{kp+k'P', \rho\right\}$, where $k$ and $k'$ are the rate constants for the singlet and triplet reaction channels, and $P$ and $P'$ are projection operators onto the singlet and triplet subspaces, respectively \citep{Haberkorn-1976, Ivanov-ea-2010}.

The evolution of $\rho$ according to eq.~(\ref{mequation}) preserves the unit trace of the density matrix. However, if chemical kinetics takes place, the trace is no longer preserved: it tends to zero \citep{Haberkorn-1976}. At any given time, the density matrix does not tend toward the equilibrium state $\rho_{\infty}$ but rather toward a reduced matrix, $\tr(\rho)\rho_{\infty}$. To account for this, we include dissipator $-g[\rho -\tr(\rho)\rho_{\infty}]$, where the reduced density matrix replaces the equilibrium one. Thus, the combined effect of thermal relaxation and chemical kinetics on the RPM process is governed by the master equation \begin{equation} \label{Lind} \partial_t \, \rho = -i \left[ {\hat H} ,\rho \right]  - \frac12 \left\{  k P + k' P', \rho \right\} - g \left[\rho- \tr (\rho) \rho_{\infty } \right] \end{equation}

If chemical kinetics is absent, that is, $k=k'=0$, then $\tr(\rho) =1$, and the equation reduces to (\ref{mequation}). A common approximation, used in \citep[e.g.,][]{Schulten-ea-1978, Zhang-ea-2014}, assumes $k'=k\neq 0$, under which the sum of the singlet and triplet projectors equals the identity operator. In this case, the chemical term in eq.~(\ref{Lind}) simplifies to $-k\rho$, leading to an exponential decay of the density matrix trace. It can be shown, that the solution to eq.~(\ref{Lind}) under this approximation is \begin{equation} \label{soluti} \rho = (U \rho_0 U^{\dagger} - \rho_{\infty})\, {\textrm{e}}^ {- (g+k)t}+ \rho_{\infty} {\textrm{e}}^ {-kt}	\end{equation} where $\rho_0$ is the density matrix of the initial state, typically an electronic singlet, fully mixed in the nuclear spin states, $\rho_0 = P/\tr(P)$, and $U(t) \equiv\exp(-i{\hat H}t)$.

Next, we examine a simple RPM scenario involving a system of two electrons and a nucleus associated with one of them. The Hamiltonian includes only the Zeeman interaction and the isotropic part of the hyperfine coupling, ${\hat H}\equiv - h\left({S}^1_z+{S}^2_z\right) + a\textbf{IS}^1$, where ${S}_i^1$ and ${S}_i^2$ are the $i$-components ($i = x, y, z$) of the spin operators for electrons 1 and 2, respectively. The constant dimensionless MF $h$ is aligned with the $z$ axis, and $\textbf{I}$ denotes the spin operator for the nucleus of electron 1. The contact hyperfine coupling constant $a$ is taken further a frequency unit.

Under the influence of MF together with chemical kinetics and spin relaxation, the ratio of the singlet $\rho_{\textrm{s}}$ and triplet populations evolves. Let $F$ denote $\tr (P U\rho_0 U^{\dagger}) $. Since the equality $\tr(P\rho_{\infty}) = 1/4$ is valid for a three-spin system, it follows from solution (\ref{soluti}) that \begin{equation} \label{eqforYS} \rho_{\textrm{s}}(t,h,k,g) \equiv \tr (P \rho ) =  [ F(t,h) - 1/4] {\textrm{e}}^{-(k+g)t} +  {\textrm{e}}^{-kt}/4 \end{equation} The magnetic effect is conventionally characterized by the time-averaged value of $\rho_{\textrm{s}} (t,h,g,k)$, where the averaging period is taken to be much longer than the timescale of rapid Torrey oscillations in $\rho_{\textrm{s}}$. For practical purposes, we define the magnetic effect $M$ such that $M=0$ in the absence of a MF, and $M>0$ in a weak MF comparable to the geomagnetic field, \begin{equation} \label{MEdef} M(h,g,k) \equiv 1- \frac{\int_0^{\infty}   \rho_{\textrm{s}} (t, h, g, k)   \mathrm{d}t} {\int_0^{\infty}   \rho_{\textrm{s}} (t, 0, g, k) \mathrm{d} t} \end{equation} Then, for example, $M=0.1$ means a 10-\% effect.

To simplify the analysis, we introduce two key variables: $\theta\equiv(g+k)^{-1}$, representing the characteristic total decay time of the density matrix, and $s\equiv k/g$, quantifying the asymmetry between the chemical kinetics and thermal relaxation rates. It turns out that these variables govern the shape and scale of the MF-dependencies, respectively. By evaluating the integrals in definition (\ref{MEdef}) using $\theta $ and $s$, one can derive an explicit functional form $M(h,\theta,s) = s \theta ^4 / ( 8 s +5 s \theta ^2  +2 \theta ^2  +2 ) A_1(h, \theta ) / A_2(h, \theta )$, where \multlinegap=0cm \begin{multline*} A_1(h, \theta ) \equiv h^2  \left( h^4 \theta ^8  -4 h^4 \theta ^6  -32 h^4 \theta ^4  +2 h^2 \theta ^8  +6 h^2 \theta ^6  -12 h^2 \theta ^4 \right. \\ \left. -64 h^2 \theta ^2  +8 \theta ^6  +16 \theta ^4  -24 \theta ^2  -32 \right), \end{multline*} \vspace{-10mm} \begin{multline*} A_2(h, \theta ) \equiv \left( h^2 \theta ^2  + \theta ^2  +1 \right)  \left( h^2 \theta ^4  +4 h^2 \theta ^2  -4 h \theta ^2  +4 \theta ^2  +4 \right) \\ \times \left( h^2 \theta ^4  +4 h^2 \theta ^2  +4 h \theta ^2  +4 \theta ^2 +4 \right). \end{multline*}

\begin{figure} [htbp] \centering \includegraphics[width=0.4\linewidth]{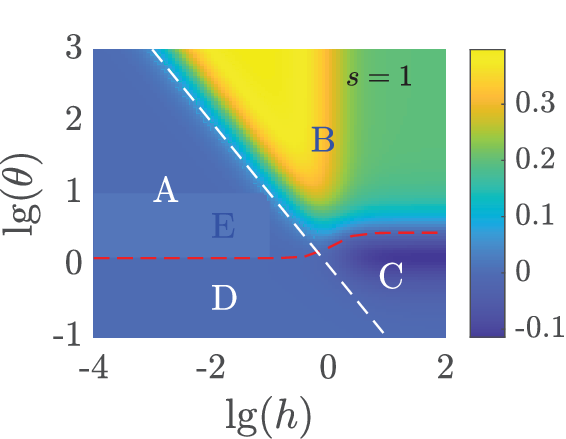} \caption{The magnitude distribution of magnetic effects $M$ as a function of parameters $h$ and $\theta$. The parameter space is divided into four distinct regions, A, B, C, D, each exhibiting qualitatively different magnetic responses. The white dashed line $h\theta=1$ separates regions with strong effects, B and C, from those with very weak effects, A and D. The red dashed line $M (h,\theta,s) =0$ demarcates regions of positive effects, A and B, from negative effects, C and D. }
\label{fig:areas} \end{figure}

The parameters $h$ and $\theta$ span distinct regions, labeled in Fig.~\ref{fig:areas} by letters, each exhibiting magnetic effects $M$ that differ either in sign or substantially in magnitude. The oblique dashed line $h\theta=1$ serves as a boundary: above and to the right of this line, the effects are relatively large, while below and to the left, they diminish by orders of magnitude.

The conversion from dimensionless variables to physical quantities in all preceding expressions is performed through the following relations, \begin{equation} \label{dimension} t \equiv a t' , ~~ h\equiv \gamma H/a, ~~ g\equiv \Gamma /a = 1/(a \tau),~~ k\equiv \kappa/a,~~ k' \equiv \kappa'/a \end{equation} where $t'$ represents time, $\gamma $ is the electron gyromagnetic ratio,  $\Gamma$ denotes the relaxation rate with $\tau$ being the relaxation time, $\kappa$ and $\kappa'$ are the chemical kinetics rates in the singlet and triplet channels, respectively. Based on the  definitions in (\ref{dimension}), the dimensional form of the relation $h\theta \sim 1$ can be obtained in the form \begin{equation} \label {gflim} \gamma H \tau \sim 1 +\kappa \tau \end{equation} For slow chemical kinetics, $\kappa\tau <1$, the condition $\gamma H \tau \sim 1$ emerges. As shown in \citep{Binhi-2025-BF}, this relationship follows from a fundamental quantum principle, the energy-time uncertainty ratio, which the RPM naturally satisfy. In the opposite regime of fast chemical kinetics, $\kappa\tau>1$, where the reaction rate exceeds spin decoherence rate, the critical magnetic field condition becomes $\gamma H \sim \kappa$, defining the threshold for observing significant magnetic effects.

In conclusion, an analytical solution (\ref{soluti}) to the master equation (\ref{Lind}) has been found above for a system consisting of two electrons and a nucleus, with the Zeeman and hyperfine interactions, taking into account quantum coherence relaxation and chemical kinetics for the case of equal rates in the singlet and triplet channels. It has been shown that the magnetic RPM effects can only be noticeable when the fundamental inequality (\ref{gflim}) is satisfied. This relationship, which is implicitly present in the RPM, is consistent with the experience of spin chemistry. If at least one of the rates of chemical processes and spin decoherence is high, the magnetic effect will not be significant.

\end{document}